\documentclass[amsmath,twocolumn]{revtex4}

\usepackage{graphicx}
\usepackage{tikz}

\usepackage{lipsum}

\begin{document}
%
\title{Non-Hermitian Acoustic Metamaterials: the role of Exceptional Points in sound absorption}
\author{V.~Achilleos}
\author{G.~Theocharis}
\author{O.~Richoux}
\author{V.~Pagneux}
\affiliation{LUNAM Universit\'e, Universit\'e du Maine, LAUM UMR CNRS 6613, Av. O. Messiaen, 72085 Le Mans, France}	
\begin{abstract}
Effective non-Hermitian Hamiltonians are obtained to describe coherent perfect 
absorbing and lasing boundary conditions. $\mathcal{PT}$-symmetry of the Hamiltonians 
enables to design configurations which perfectly absorb at multiple frequencies.
Broadened and flat perfect absorption is predicted at the exceptional point of $\mathcal{PT}$-symmetry breaking
while, for a particular case, absorption is enhanced with the use of gain.
The aforementioned phenomena are illustrated for acoustic scattering through Helmholtz resonators
revealing how tailoring the non-Hermiticity of acoustic metamaterials leads to novel mechanisms for enhanced absorption.
\end{abstract}

\maketitle

\section{Introduction}

In both classical and quantum physics, non-Hermitian Hamiltonians are
being used to describe open systems~\cite{moiseyev} and systems featuring 
energy dissipation and/or gain~\cite{Heiss,Cao}. 
Most of the interesting phenomena in non-Hermitian systems, manifest close to the
exceptional points (EPs) which are spectral singularities appearing when two eigenmodes coalesce~\cite{Heiss,Cao,Rotter1}.
 A particular class of non-Hermitian systems, featuring balanced gain and loss,  are called 
$\mathcal{PT}$ -symmetric~\cite{Bender98}, and have attracted considerable attention. 
Such systems suggest possible generalization of quantum mechanics and are extensively studied in diverse areas 
of physics including Bose-Einstein condensates \cite{Cartarius,Achilleos}, opto-mechanics \cite{Lu}, acoustics~\cite{Fleury,Shi}
and mostly in optics~\cite{Ruter,Ramezani,Regensburger,Peng14a,Peng14b,Chang,Feng14,Hodaei,Jing}.
 Operating around the EPs of $\mathcal{PT}$-symmetric scattering systems,
has lead to the observation of extraordinary wave effects such as unidirectional invisibility~\cite{Mostafazadeh,Feng12}
optical isolation~\cite{Chang}, unidirectional perfect absorption~\cite{Ramezani16},
the simultaneous appearance of coherent perfect absorption (CPA) and lasing \cite{Longhi,Chong11}
or non-reciprocal light propagation~\cite{Peng14a}. 

Beyond $\mathcal{PT}$-symmetric systems, non-Hermitian configurations featuring unequal amount of loss and gain 
(or only losses) also exhibit interesting phenomena, closely related with the existence of EPs.
These include CPA~\cite{Chong10}, unidirectional near total light absorption~\cite{Veronis1},
chiral modes and directional lasing~\cite{Rotterpnas} and mode switching~\cite{Rotternature}.
In this context, effective non-hermitian Hamiltonians associated with the scattering matrix
have been obtained to describe CPA, both in microwave~\cite{Sun} and polaritonic~\cite{Zanotto14,Zanotto16}
systems. 

Especially in acoustics, the phenomenon of perfect absorption has attracted great attention the 
last years due to its direct applications to numerous noise reduction problems. 
Many solutions have been proposed in the low frequency regime based on subwavelength metamaterial designs,
by critically coupling~\cite{Xu00} resonant scatterers to the waveguide  i.e. by balancing the energy leackage and the internal
losses of the resonances. Such studies include the use of membranes~\cite{Ma,Wei,Romerosr},
the concept of slow sound~\cite{Groby1,Groby2} and Helmholtz resonators both in the linear~\cite{Merkel,Romero16,Noe} and nonlinear~\cite{Achilleos16} regimes.  

Here, we show that by properly tuning the non-Hermiticity of acoustic metamaterials and exploiting the 
appearance of EPs is a novel way to control sound absorption.
The non-Hermiticity of acoustic metamaterials has only been exploited very recently.
Specifically, in Ref.~\cite{PRXacoustics} a closed system of resonant 
cavities was used to study higher order EPs, in Ref.~\cite{nonhermitianacoustic}
an open lossy acoustic system is used to achieve high quality acoustic hologram,
and in Ref.~\cite{bi} the mode coupling around EPs is studied for a lined waveguide.
In this work we derive, directly from the acoustic equations, a coupled mode theory (CMT) for the scattering of low frequency waves, 
in a waveguide side loaded by $N$ Helmholtz Resonators (HR) at the same position. 
In Section 2, we employ the CMT and obtain two different effective Hamiltonians 
corresponding to  different boundary conditions of the scattering problem : no outgoing waves (for CPA)  and no
incident waves (for laser). For no outgoing waves, the eigenvalues of the effective Hamiltonian yield the particular
frequencies where CPA is obtained, which are generally complex since the matrix is non-Hermitian.
For the case of two HRs we show how the avoided crossings around an EP lead to the appearance of CPA for real frequencies. 
In Section 3, $\mathcal{PT}$-symmetry of the effective Hamiltonian is found to provide the necessary condition for 
multiple CPAs. For the case of two resonators the conditions for 
$\mathcal{PT}$-symmetry coincide with the critical coupling (CC) of each resonator with the waveguide. 
Surprisingly, in the case of three HRs, critical coupling of each resonator does not lead to multiple CPAs. 
$\mathcal{PT}$-symmetry however can be established for three resonators, leading to three perfectly absorbed frequencies obtained 
by the aid of acoustic gain. At the EPs separating the $\mathcal{PT}$-unbroken and broken phases, we observe
flattened absorption, which  is analytically approximated and is found to be stronger for higher order EPs.
\begin{figure}[htb]
\includegraphics[width=8.5cm]{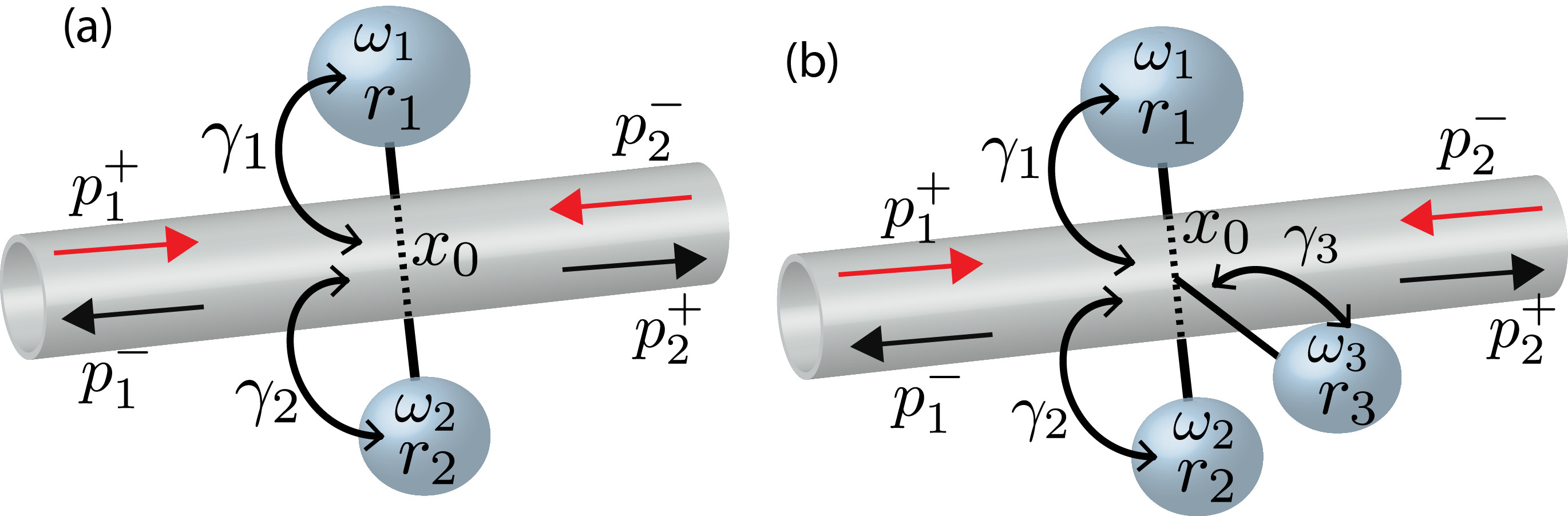}
\caption{A schematic illustration of the configurations under consideration, using two (dimer) and three (trimer) point resonant scatterers side loaded to a waveguide. 
 $\omega_{1,2,3}$ are the resonant frequencies,  $r_{1,2,3}$ the intrinsic losses and $\gamma_{1,2,3}$ the coupling strengths. Arrows
depict the incoming and outgoing waves in the waveguide.}
\label{sketch_gen} 
\end{figure}
%

%
%

\section{Non-Hermitian effective Hamiltonians}

\subsection{Coupled mode theory for $N$ resonators}
We consider the scattering of plane waves by $N$ resonant scatterers, side loaded to a waveguide 
at $x_0$ [see Fig.~(\ref{sketch_gen})], which without loss of generality we assume to be $x_0=0$. 
The subwavelength resonators, which can be considered as point scatterers, are characterized by a resonance frequency $\omega_i$, 
internal losses $R_i$ and external coupling strengths $\Gamma_i$ ($i=1,2\ldots N$). 
The conventional way to study the system, is by using the scattering matrix $\mathbf{S}$,
which relates the incident waves $p_1^+$ and $p_2^-$ with the outgoing waves
$p_1^-$ and $p_2^+$, given by
\begin{eqnarray}
\left(
\begin{array}{c} p_2^+\\p_1^-\end{array} \right)=\mathbf{S}(\omega)\left(
\begin{array}{c} p_1^+\\p_2^-\end{array} \right) \equiv
\left( \begin{array}{cc} t&r_R \\ r_L&t 
\end{array}\right)
\left(\begin{array}{c} p_1^+\\p_2^-\end{array} \right).
\end{eqnarray}
$r_L$, and $ r_R$ are the complex reflection coefficients 
for left and right incidence and $t$ is the complex transmission coefficient of the reciprocal system.
For the point symmetric scatterers considered here, $r_L=r_R=r$ and  $t=1+r$~\cite{Merkel}. 
Using these relations, one can show that the determinant of the scattering
matrix is given by
\begin{eqnarray}
{\rm det} (\mathbf{S})=t+r.
\label{dets}
\end{eqnarray}


Following a different approach which we find convenient, the scattering properties can also be obtained using
the continuity of the field and the conservation of flux in the waveguide, along with the evolution of the field
in the resonant scatterers. The corresponding CMT is derived in the Appendix A, and the resulting equations are 
\begin{align}
&p_1^{+}(t)+p_1^{-}(t)=p_2^{+}(t)+p_2^{-}(t),
\label{cont12}\\
&p_1^+(t)-p_1^-(t)=p_2^+(t)-p_2^-(t)+\sum_i 2\frac{\gamma_i}{\Omega_i^2}\dot{p}_{ci},\label{cont22} \\
&\ddot{\vec{P}}+\mathbf{R}\dot{\vec{P}}+\mathbf{K}\vec{P}=\vec{F},
\label{kyrillov}
\end{align}
where $(\dot{})$ denotes differentiation with respect to the normalized time $\textit{t}\rightarrow\omega_1\textit{t}$.
In Eq.~(\ref{kyrillov}), $\vec{P}=[p_{c1},\ldots, p_{cN}]^T$ is the field amplitude in the $N$ resonators, and  $\vec{F}=[\Omega_1^2(p_1^++p_2^-),\ldots,\Omega_N^2(p_1^++p_2^-)]$ is the external driving force due to the incoming waves from the waveguide. The matrix $\mathbf{K}$ is diagonal and its
elements are the normalized resonant frequencies $\Omega_i^2$. The matrix $\mathbf{R}$  includes the internal losses ($r_i$),
and the leakage of the resonators to the waveguide  ($\gamma_i$).
 Below using Eqs.~(\ref{cont12})-(\ref{kyrillov}), we derive effective Hamiltonians 
to describe specific boundary conditions of the scattering problem.

\paragraph*{Perfect absorption.---\!\!\!}

This boundary condition of special interest is the one of \textit{no outgoing waves}
$p_1^{-}=p_2^{+}=0$, which when used in Eq.~(\ref{cont12}) leads to 
\begin{equation}
p_1^{+}=p_2^{-}.
\label{coherent}
\end{equation}
Eq.~(\ref{coherent}) shows that, in a two port system with point scatterers side loaded at the same position,
perfect absorption occurs only when the incoming waves have equal amplitudes \textit{and} phase.
Under these conditions, we may re-write Eq.~(\ref{kyrillov}) in the usual 
Hamiltonian matrix form 
\begin{align}
\dot{i\vec{\Psi}}=\mathbf{L}\vec{\Psi},
\label{vecode1}
\end{align}
where  $\vec{\Psi}=[p_{ci},\ldots, p_{cN},\dot{p}_{c1},\ldots,\dot{p}_{cN}]^T$.
%
Assuming  harmonic solutions of Eq.~(\ref{vecode1}) in the form $\vec{\Psi}(t)=e^{-i\omega t}\vec{\Psi}_0$,
we obtain the following eigenvalue problem
\begin{eqnarray}
(\mathbf{L}-\omega\mathbf{I})\vec{\Psi}_0=0.
\label{eig1}
\end{eqnarray}
In general the eigenvalues of the non-Hermitian matrix $\mathbf{L}$ are \textit{complex}.
When an  eigenvalue becomes real, at that particular frequency the in-phase incoming
waves will be completely absorbed and this corresponds to a CPA.

\paragraph*{Laser. ---\!\!\!}
Another boundary condition of interest, is the one of \textit{ no incoming waves}, $p_1^{+}=p_2^{-}=0$
leading to $\vec{F}=0$.
For this boundary condition, we can also write the corresponding
system of equations and the eigenvalue problem as
%
\begin{align}
\dot{i\vec{\Psi}}=\mathbf{H}\vec{\Psi},\quad (\mathbf{H}-\omega\mathbf{I})\vec{\Psi}_0=0.
\label{vecode2}
\end{align}
The non-Hermitian matrix $\mathbf{H}$ can be obtained from $\mathbf{L}$ using the relation
\begin{align}
&\mathbf{H}(r_i,\gamma_i,\Omega_i) =\mathbf{L}(r_i,-\gamma_i,\Omega_i).
\label{laserbc}
\end{align}
The complex eigenvalues of matrix $\mathbf{H}$ correspond to the \textit{quasibound} states of the system. 
Their real part corresponds to the resonance frequency and their imaginary part describes the width of the resonance.
When an eigenvalue of $\mathbf{H}$ becomes real, the system appears to have 
a finite output without any input (since we chosen the boundary $p_1^{+}=p_2^{-}=0$ for $\mathbf{H}$ ),
and this is a \textit{lasing}  frequency of the system. Eq.~(\ref{laserbc}) describes in a compact way the connection
between the two sets of boundary conditions:  laser frequencies described by $\mathbf{H}$  have a CPA partner
which corresponds to changing the boundary conditions and the sign of the leakage.
 
A direct connection between the scattering matrix 
and the effective Hamiltonians can be established~\cite{Zanotto16} by noting that 
the determinant of $\mathbf{S}$ can be written as:
\begin{eqnarray}
{\rm det}(\mathbf{S})=-\frac{{\rm det}(\mathbf{L}-\omega\mathbf{I})}{{\rm det}(\mathbf{H}-\omega\mathbf{I})}.
\label{conne}
\end{eqnarray}
Eq.~(\ref{conne})  shows that
the eigenvalues of $\mathbf{L}$ are the\textit{ zeros} of the determinant 
of the scattering matrix, and the eigenvalues of $\mathbf{H}$ are its \textit{poles}.
%
\subsection{Exceptional points of the effective Hamiltonians}
\begin{figure}[tbp!]
\includegraphics[width=8.5cm]{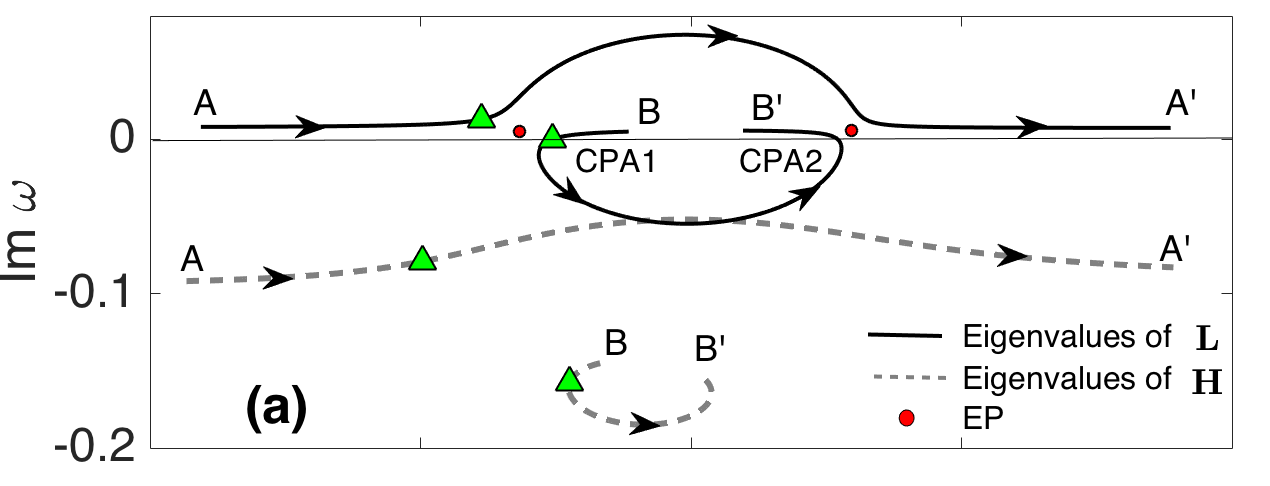}
\includegraphics[width=8.5cm]{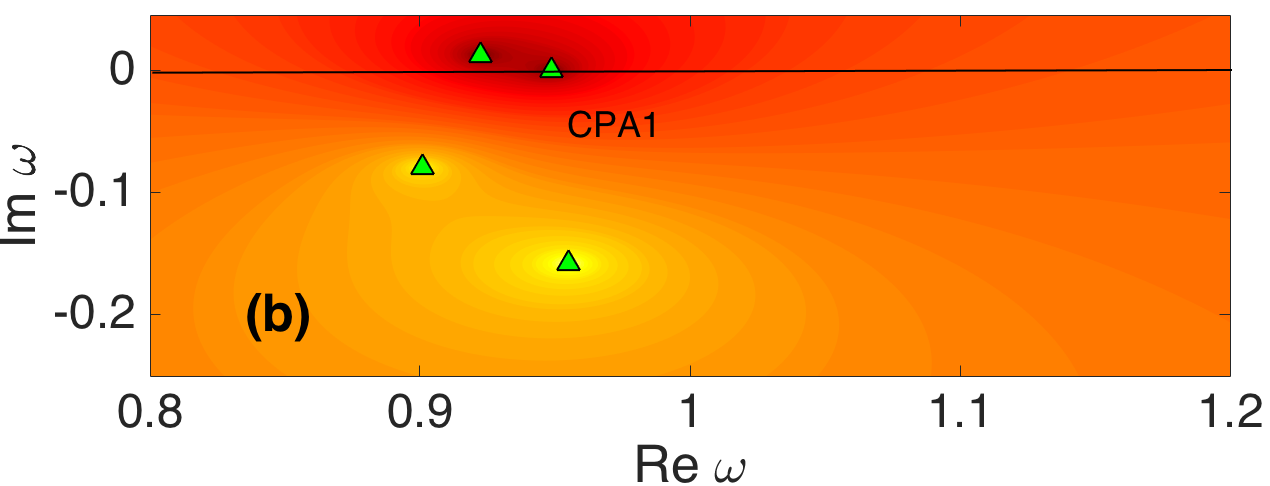}
\caption{(a) Solid (dashed) lines show the eigenvalues of  $\mathbf{L}$ ($\mathbf{H}$) for the dimer
at parameter values $\gamma_1=0.15$, $\gamma_2=0.1$, $r_1=0.1386$ and $r_2=0.085$ while
the running parameter is $\Omega\in [0.8,1.2]$.  Triangles indicate the eigenvalues corresponding to CPA1.
(b)  Determinant of $\mathbf{S}$ in the complex frequency plane for parameters corresponding to
the CPA1 of panel (a) }
%
\label{fig2} 
\end{figure}
In this work, we focus on  two cases: the dimer  with two, and the trimer with three subwavelength resonators,
as shown in panels (a) and (b) of Fig.~\ref{sketch_gen} respectively. 
The exact form of the effective Hamiltonians $\mathbf{L}$  for both the  dimer  and trimer 
are given by  Eq.~(\ref{4x4_1}) and Eq.~(\ref{4x4_3}) in the appendix respectively.
The eigenvalues $\omega$ of the matrices  $\mathbf{L}$ can be found from the corresponding
secular equations
%
\begin{align}
&f_{L2}=\omega^4-i\alpha\omega^3+\beta\omega^2+ i\mu\omega+\nu =0,
\label{secular} \\
&f_{L3}=\omega^6-i\lambda'\omega^5+y'\omega^4+i\alpha'\omega^3+\beta'\omega^2+ i\mu'\omega+\nu' =0,
\label{secular3}
\end{align}
for the dimer and trimer respectively. The polynomials $f_{L2}(\omega,r_i,\gamma_i,\Omega_i)$ and
$f_{L3}(\omega,r_i,\gamma_i,\Omega_i)$ are functions of $\omega$ and depend on  the $3N-1$ parameters
$r_i$, $\gamma_i$ and $\Omega_i$ (since $\Omega_1=1$).
 The corresponding equations for the matrices $\mathbf{H}$ are given by
\begin{align}
f_{H2}=f_{L2}(\omega,r_i,-\gamma_i,\Omega_i), \\
f_{H3}=f_{L3}(\omega,r_i,-\gamma_i,\Omega_i),
\end{align}
due to Eq.~(\ref{laserbc}).
Note that the polynomials also satisfy the relation 
$f_H(\omega^*,-r_i,\gamma_i,\Omega)= f_L(\omega,r_i,\gamma_i,\Omega)$. 
For the lossless case ($r_i=0$) the latter relation recovers the well known fact, that the
zeros and poles of the scattering matrix are complex conjugates due to time reversal symmetry.

An appealing property of non-Hermitian matrices, is the existence of EPs and the associated wave phenomena around them. 
Here, we investigate the wave scattering around the EPs of the non-Hermitian matrix $\mathbf{L}$
and their connection with the emergence of CPA.
To start with, we consider the dimer. The parametric space of the relevant  matrix $\mathbf{L}$ 
is 5-dimensional ($r_{1,2},\gamma_{1,2},\Omega_2$)  and we simplify the problem by reducing it to the 2-dimensional space of $\Omega_2$ and $r_2$.
The coupling coefficients  $\gamma_{1,2}$  are fixed to the moderate values of $\gamma_1=0.15$ and $\gamma_2=0.1$.
These values correspond to HRs which are of  moderate Q factor (Q$\sim 50$)
in the low frequency regime. Additionally, the losses of one of the HRs  $r_1=0.138$ are set to a relatively high value,
since we have in mind highly absorbing structures. 

As it is shown in Appendix C, we locate two EPs  in the two dimensional parameter  space
($\Omega_2,r_2$). These are found at  $\omega_{\rm EP1}\approx 0.9367+ i 0.005$
and $\omega_{\rm EP2}\approx 1.0595+ i 0.006$.
For paremeter values close to the EPs, the complex eigenvalues are 
enforced to repel each other through an avoided crossing.
An example is shown in Fig 2 (a), where the trajectories of the two complex eigenvalues of $\mathbf{L}$ are plotted
with the solid lines for $\Omega_2\in [0.8,1.2]$ and $r_2=0.086$.
%
%
The two eigenvalues, initially located at points A and B for $\Omega=0.8$,  as
they approach the leftmost EP [leftmost (red) circle in panel (a)]  they repel, and one crosses 	 the real axis.
Exactly at the point where this eigenvalue becomes real a CPA emerges.
By further increasing $\Omega_2$, the eigenvalue crosses the real axis around the rightmost 
EP, signaling a second CPA, and ends at point B$'$ when $\Omega=1.2$.
The dashed lines in Fig.~\ref{fig2} (a), depict the corresponding complex eigenvalues of $\mathbf{H}$
(solutions of $f_{H2}$). Note that the eigenvalues do not coalesce rather than their trajectories 
meet at some point.
%

Figure~\ref{fig2} (b) is presented to 
visualize the relation between the eigenvalues of $\mathbf{L}$ and $\mathbf{H}$ and the zeros and poles
 of  ${\rm det} (\mathbf{S})$. In particular we show  ${\rm det} \mathbf({S})$ in the complex frequency 
 plane for the parameters corresponding to CPA1 of panel (a). The (green) triangles indicate the 
 %
respective eigenvalues from panel (a).

\section{$\mathcal{PT}$-symmetric effective Hamiltonians}
In the previous Section it was shown that, a CPA frequency appears when an 
eigenvalue of $\mathbf{L}$, around an EP, is driven across the real axis and becomes real.
It is known however, that when a non-Hermitian matrix is $\mathcal{PT}$-symmetric,
it is possible to obtain regions where\textit{ all} eigenvalues are real. It is very interesting 
then that $\mathcal{PT}$-symmetry provides a condition to obtain multiple CPA
frequencies. Also, it
ensures the emergence of EPs on the real axis, which as we will show below, leads
to interesting results regarding broadened flat perfect absorption.		

It appears that  $\mathcal{PT}$-symmetry is enforced quite easily:
we simply require the corresponding secular equations of the non-Hermitian
matrix $\mathbf{L}$   to be real~\cite{Bender2010}.
Then, the eigenvalues are either real or complex conjugates.
%
%
\begin{figure}[tbp!]
\includegraphics[width=8.5cm]{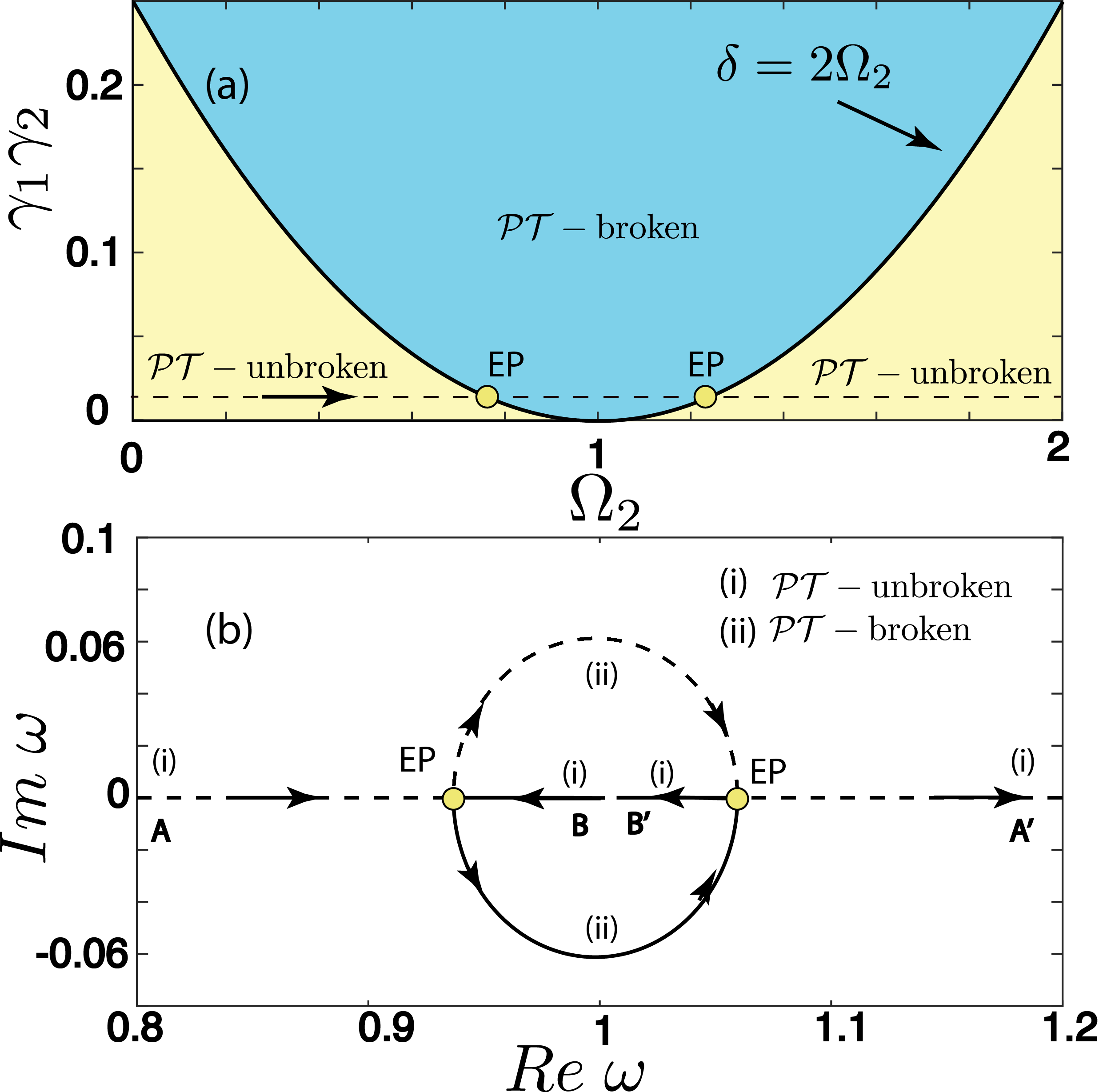}
\caption{(a) The phase plane indicating the $\mathcal{PT}$-broken and unbroken phases when both
resonators are critically coupled. The solid line parabola indicates the EPs separating the two phases.
(b) A trajectory of the eigenvalues of matrix $\mathbf{L}$  corresponding to the dashed horizontal line
of panel (a) with $\gamma_1=r_1=0.15$,  $\gamma_2=r_2=0.1$,
and $ \Omega_2\in [ 0.8, 1.2] $. Arrows indicate the direction of the eigenvalues as $\Omega_2$ increases.
EPs points are at $\Omega_2=1\pm \sqrt{\gamma_1\gamma_2}$.}
\label{cont_zeros} 
\end{figure}

\subsection{Dimer}
From Eq.~(\ref{secular}) we immediately obtain that  $\mathcal{PT}$-symmetry requires 
$\alpha=\mu=0$ leading to the following equations
\begin{align}
&r_1-\gamma_1+r_2-\gamma_2=0, \label{cond1} \\
&\Omega_2^2(r_1-\gamma_1)+r_2-\gamma_2=0.\label{cond2}
\end{align}
According to Eqs.~(\ref{cond1})-(\ref{cond2}), two different $\mathcal{PT}$ configurations are possible:
(i)  $\Omega_2=1$ with the  rest of the parameters satisfying Eq.~(\ref{cond1}), and  (ii)
each of the resonator is critically coupled (i.e. $r_i=\gamma_i$) and the frequency $\Omega_2$ is freely chosen.
 In the former case, the system is always found  to be in the $\mathcal{PT}$-broken phase and the eigenvalues of 
$\mathbf{L}$ are always complex conjugates exhibiting no CPA. We thus focus our study in case (ii) 
where each resonator is critically coupled.  This is an intriguing generalization of the trivial case 
with one resonance. It is known that the  CPA condition for a single resonator is satisfied
for a critically coupled resonance with $r_1=\gamma_1$~\cite{Wei}. Here we have found that by adding 
another critically coupled resonator we can obtain two CPA frequencies. However as we will
show below, this pattern is not valid for more resonators since the interactions between the 
eigenvalues become more complex.

Choosing $r_1=\gamma_1$ and $r_2=\gamma_2$ , we can rewrite the secular equation~(\ref{secular}) as
\begin{eqnarray}
f_L=(\omega-\tilde{\omega}_{1})(\omega+\tilde{\omega}_{1})(\omega-\tilde{\omega}_{2})(\omega+\tilde{\omega}_{2}),
\label{solutions}
\end{eqnarray}
where, $\tilde{\omega}_{1,2}$ are the eigenvalues of $\mathbf{L}$ given by
\begin{eqnarray}
\tilde{\omega}_{1,2}=\frac{1}{\sqrt{2}}\sqrt{\delta\pm \sqrt{\delta^2-4\Omega_2^2}},
\end{eqnarray}
with $\delta=1+\Omega_2^2+\gamma_1\gamma_2$.
In Fig.~\ref{cont_zeros} (a) we show a phase diagram,
 indicating the broken and unbroken $\mathcal{PT}$-symmetric phases. 
In the unbroken phase [outside the parabola (yellow)] where  $\delta>2\Omega_2$, the eigenvalues $\tilde{\omega}_{1,2}$ 
are real and correspond to two different simple CPA frequencies of a particular configuration. 
On the contrary, in the broken phase [inside the parabola (blue)], when $\delta<2\Omega_2$,
 the eigenvalues $\tilde{\omega}_{1,2}$  are complex conjugates and the system features no CPA. 
The solid black line of Fig.~\ref{cont_zeros} corresponds to $\delta=2\Omega_2$ and indicates the family of
EPs separating the two phases, with real eigenvalues $\tilde{\omega}_{1}=\tilde{\omega}_{2}\equiv\omega_{\rm 
EP}$. The EPs correspond to the coalescence of two CPA frequencies which we, further-on, call
\textit{double} CPA. 

In Fig.~\ref{cont_zeros}~(b), we plot the eigenvalues of  $\mathbf{L}$ for the trajectory indicated
by the horizontal dashed line of panel (a).
Initially the system is in the unbroken phase with  two different real eigenvalues corresponding to two CPA 
frequencies (points A and B). As $\Omega_2$ is increased, they approach each other along the real axis and collide at an EP. Note
that in contrast with the non $\mathcal{PT}$-symmetric case shown in Fig.~\ref{fig2} (a), the EP is now on the real axis.
Beyond this EP the $\mathcal{PT}$-symmetry is broken and the system does not have a CPA. 
As $\Omega_2$ increases, a second EP appears where the two eigenvalues coalesce, and become real again 
re-entering the unbroken phase.
Note the differences of Fig.~\ref{cont_zeros}~(b) with the solid lines in Fig.~\ref{fig2} (a) and especially
the fact that the two EPs have been shifted on to the real axis.

%
\begin{figure}[tbp!]
\includegraphics[width=8.7cm]{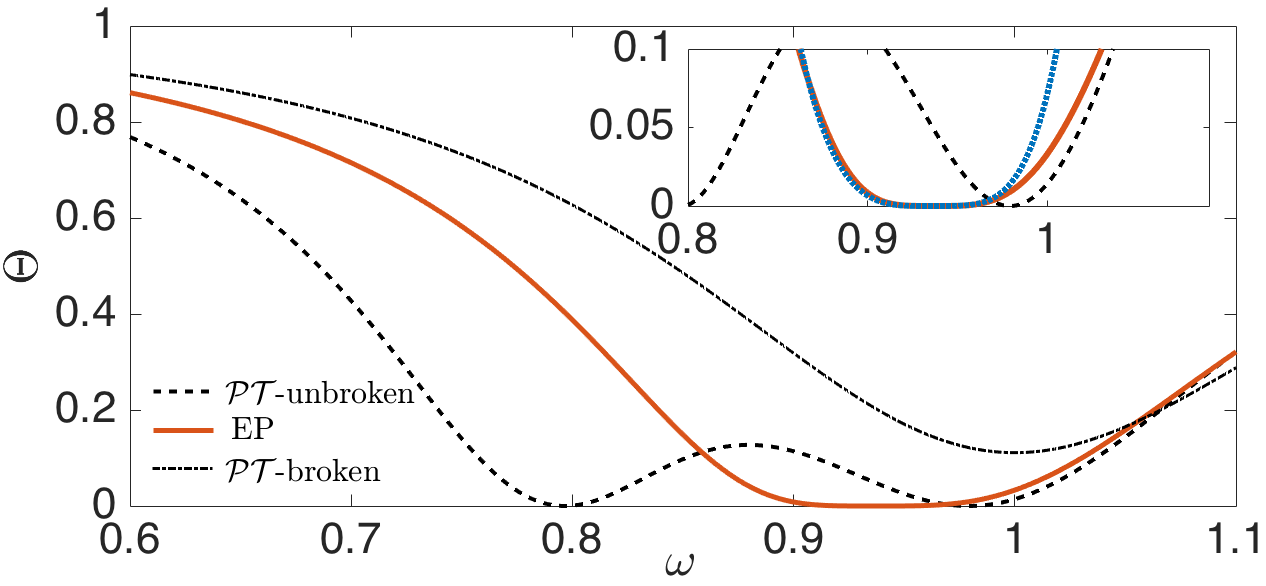}
\caption{The parameter $\Theta$ as a function of frequency for the same parameters as in Fig.~\ref{cont_zeros} (b).
The solid (red) line corresponds to the leftmost EP, the dashed (black) in the $\mathcal{PT}$-unbroken  ($\Omega_2=1.2$)  and the dashed-dotted (black) line in $\mathcal{PT}$-broken  ($\Omega_2=1$) regions.  The inset shows a
close up of $\Theta$ to illustrate the difference between a standard CPA and the CPA at the exceptional point. The blue dotted
line indicates the analytical approximation of Eq.~(\ref{thetaapprox}).
 }
\label{theta_conf1} 
\end{figure}
The next step is to illustrate what is the influence of the proposed
 $PT$-symmetric configurations in the scattering properties of he system. 
We first note that the occurrence of CPA and laser frequencies can both be captured using 
the ratio between outgoing and incoming field intensity $\Theta$ defined as
\begin{eqnarray}
\Theta=\frac{|p_1^-|^2+|p_2^+|^2}{|p_1^+|^2+|p_2^-|^2},
\label{theta}
\end{eqnarray}
%
where $\Theta\rightarrow 0$ at a CPA frequency while $\Theta >1$ signals the amplification of incoming waves.
Note that in a conservative system $\Theta=1$.
To achieve CPA we are obliged to satisfy Eq.~(\ref{coherent}) by considering scattering with incoming waves of equal amplitude and phase. In that case Eq.~(\ref{theta}), can be written as 
\begin{eqnarray}
\Theta=\frac{|p_1^-|^2+|p_2^+|^2}{2|p_1^+|^2}=|t+r|^2,
\label{theta2}
\end{eqnarray}
which using Eqs.~(\ref{dets}) and (\ref{conne}) is explicitly written as
%
\begin{eqnarray}
\Theta=
\frac{|f_L(\omega,r_i,\gamma_i,\Omega_i)|^2}{|f_H(\omega,r_i,\gamma_i,\Omega_i)|^2}.
\label{theta2}
\end{eqnarray}
In Fig.~\ref{theta_conf1}, we plot $\Theta$ as a function of frequency when matrix $\mathbf{L}$ 
is $\mathcal{PT}$-symmetric. We show three different cases when $\mathbf{L}$ is in the
PT-broken phase [dashed-dotted (black) line], 
in the PT-unbroken phase [dashed (black) line] and at an EP [solid (red) line].
In the broken phase the parameter $\Theta$ is nonzero for all frequencies (no CPA).
In the unbroken phase, $\Theta=0$ at the two real eigenvalues of  $\mathbf{L}$ showing two  simple CPA.
According to Eq.~(\ref{solutions}) and Eq.~(\ref{theta2}), around each CPA frequency the parameter $\Theta$ is analogous to
 $\Theta\sim| (\omega-\tilde{\omega}_{1,2})^2|$ which is quadratic in frequency.
For the case of an EP, the coalescence of the two eigenvalues of $\mathbf{L}$  gives rise to a double CPA point 
around  $\omega=\omega _{\rm EP}$  (solid red line of Fig.~\ref{theta_conf1}).

Around the EP, Eq.~(\ref{theta2}) can be approximated by
\begin{eqnarray}
\Theta\approx \frac{|(\omega-\omega_{\rm EP})^4(\omega+\omega_{\rm EP})^4|}{|f_H(\omega_{\rm EP})|^2},
\label{thetaapprox}
\end{eqnarray}
which is plotted in the inset of Fig.~\ref{theta_conf1} by the dotted (blue) line. 
The consequence of Eq.~(\ref{thetaapprox}), which exhibits a\textit{ quartic} dependence of  $\Theta$ with frequency,
is that perfect absorption is flattened, in comparison with the simple CPA. On the other hand the denominator which depends 
on the polynomial $f_H$ controls how broadband is the perfect absorption.
%

\subsection{Trimer}
We now proceed with the case of the trimer which can display not only three CPAs
but also higher order EPs. First we identify the parameter space where
the relevant matrix $\mathbf{L}$  is $\mathcal{PT}$-symmetric.
The corresponding secular equation for the trimer is given by Eq.~(\ref{secular3})
and $\mathcal{PT}$-symmetry is enforced when $\lambda'=\alpha'=\mu'=0$,
leading to the following  conditions
\begin{align}
&r_1-\gamma_1+r_2-\gamma_2+r_3-\gamma_3=0, \label{cond13} \\
&\Omega_2^2(r_3-\gamma_3)+\Omega_3^2(r_2-\gamma_2)+\Omega_2^2\Omega_3^2(r_1-\gamma_1)=0,\label{cond33}\\
&r_2-\gamma_2+r_3-\gamma_3+\Omega_2^2(r_1-\gamma_1+r_3-\gamma_3)\nonumber \\
&+\Omega_3^2(r_1-\gamma_1+r_2-\gamma_1)+r_2r_3(r_1-\gamma_1)\nonumber \\
&-r_1(r_2\gamma_3+r_3\gamma_2)=0.
\label{cond23}
\end{align}
%
%
A surprising result of Eqs.~(\ref{cond13})-(\ref{cond23}) is the following: 
by critically coupling each of the three resonators (i.e. $r_i=\gamma_i$) we \textit{do not} obtain 
a $\mathcal{PT}$-symmetric configuration and cannot have three CPA frequencies. 
Indeed, choosing $r_i=\gamma_i$, Eqs.~(\ref{cond13}) and (\ref{cond33}) are satisfied 
while Eq.~(\ref{cond23}) is not, due to a non-vanishing term $\sim r_1(r_2\gamma_3+r_3\gamma_2)$. 
This is a rather counter-intuitive result, since this term does not vanish even when the resonant frequencies
are far away from each other.

Configurations which support three CPAs and also feature higher order EPs, 
can be obtained when Eqs.~(\ref{cond13})-(\ref{cond23}) are satisfied, and $\mathbf{L}$ is $\mathcal{PT}$-symmetric.
In this case, the eight dimensional parameter space of the trimer is reduced to a five dimensional one, which 
exhibits a plethora of regions either in the broken or unbroken phase. We choose to study a particular parametric
region in order to illustrate both the transition from the broken to the unbroken phase, and also show the
existence of a triple EP. In particular, we chose $\Omega_2=\Omega_3\equiv\Omega$ as  a varying
parameter and fix $\gamma_{1,2,3}$ and $r_1$.
%
The rest 
of the parameters ($r_2$ and $r_3$) are given by  Eqs.~(\ref{cond13})-(\ref{cond23}).
Note that this choice of parameters leads to negative values for $r_3$ namely the inclusion of acoustic gain.
Mechanisms for gain in acoustics have been realized experimentally for example in Ref.~\cite{Fleury} by the 
use of loudspeakers with tailored electrical circuits.
%
%
\begin{figure}[tbp!]
\includegraphics[width=8.5cm]{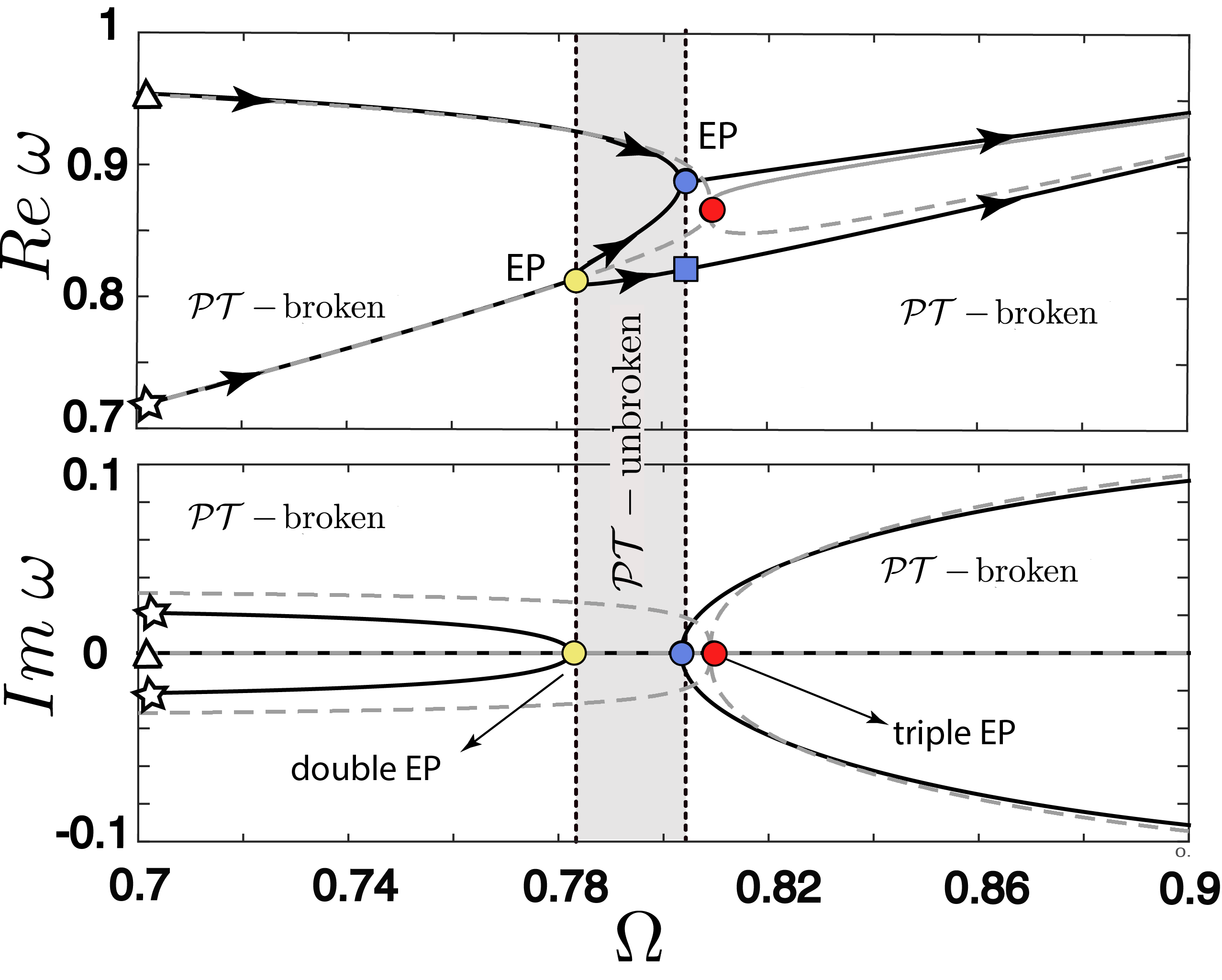}
\caption{Trajectories of the eigenvalues of  $\mathbf{L}$ given by Eq.~(\ref{secular3})
with varying parameter  $\Omega\in [0.7,0.9]$.
The rest parameters chosen are $\gamma_1=r_1=0.2$, $\gamma_2=0.2$ and $\gamma_3=0.01$.
The losses $r_2$ and $r_3$ are then given by Eqs.~(\ref{cond13})-(\ref{cond23}) as function of $\Omega$. 
Shaded region depicts the $\mathcal{PT}$-unbroken regime. 
The dashed line shows the eigenvalues for the same parameters but with $\gamma_3=0.0157$, when the system exhibits
a triple EP. }
\label{traj3} 
\end{figure}

\begin{figure}[tbp!]
\includegraphics[width=8.5cm]{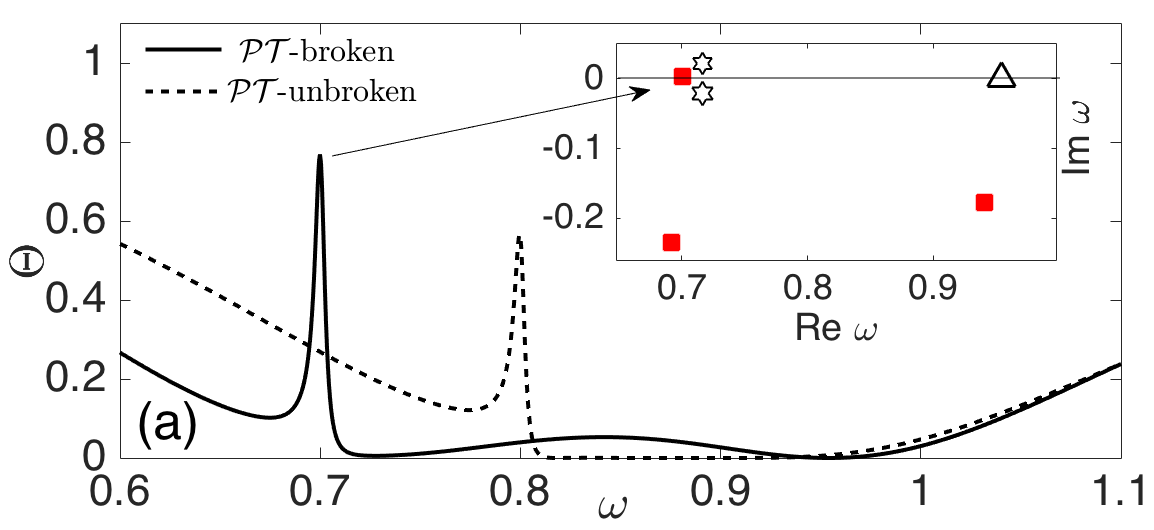}
\includegraphics[width=8.5cm]{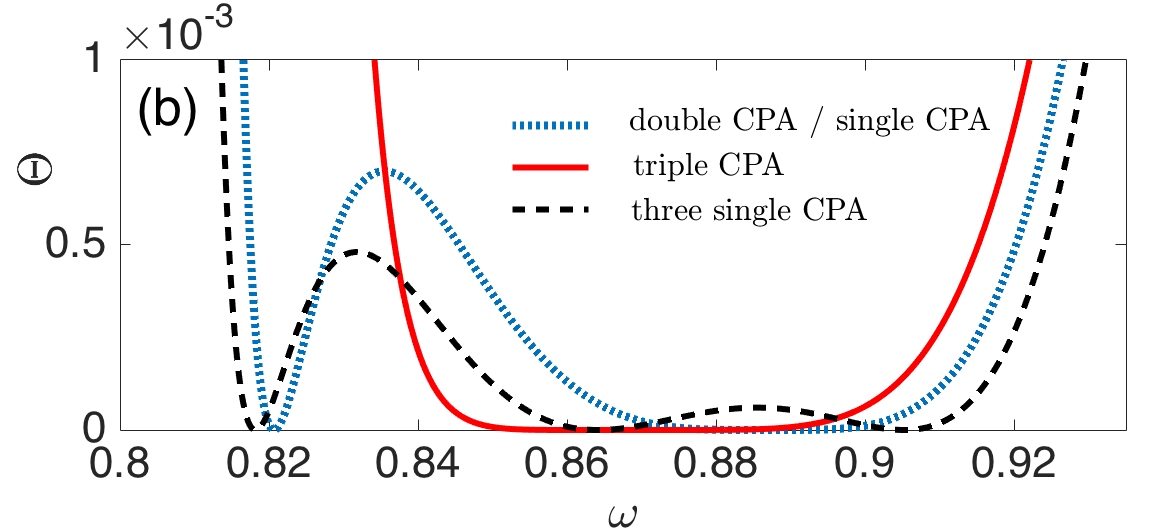}
\caption{Panel (a): Same as in Fig.~\ref{theta_conf1} for the trimer. Solid line depicts the 
 $\mathcal{PT}$-broken phase  for $\Omega=0.7$. The dashed line corresponds to a $\mathcal{PT}$-unbroken
case for $\Omega=0.78$. The inset depicts the eigenvalues of $\mathbf{L}$ (triangle and stars) and of $\mathbf{H}$
(squares). Panel (b): Dashes  line is the same is in panel (a). A zoom at the frequencies between $\omega=0.8$ and $\omega= 1$. The 
dotted (blue) line depicts the case of $\Omega=0.803$ corresponding to a simple EP (rightmost vertical line
of Fig.~(\ref{traj3})]. Solid (red) line corresponds to the triple EP with $\Omega=0.803$ and
$\gamma_3=0.0157$.}
\label{theta3} 
\end{figure}

A $\mathcal{PT}$-symmetric trajectory of the eigenvalues of $\mathbf{L}$
is shown in Fig.~\ref{traj3}.  
At the beginning of the trajectory at $\Omega=0.7$, $\mathbf{L}$ is in the $\mathcal{PT}$-broken phase
where there is one real eigenvalue (denoted by triangles) and a complex conjugate pair (denoted by stars).
Thus the corresponding curve of $\Theta$ as a function of frequency, shown with the solid line 
in Fig.~\ref{theta3} (a), exhibits one single CPA frequency at $\omega\approx 0.95$. The additional
dip around $\omega\approx 0.72$ corresponds to the complex conjugate pair indicated by the
stars in Fig.~\ref{traj3} and the inset of  Fig.~\ref{theta3} (a). The peak appearing 
in Fig.~\ref{theta3} (a) emerges due to one of the solutions of $f_{H3}$ (i.e. a pole) which 
is very close to the real axis.
%
At $\Omega=0.781$, two complex conjugate eigenvalues coalesce and become real. 
Beyond this EP,  $\mathbf{L}$ is in the $\mathcal{PT}$-unbroken phase indicated by the shaded area
in Fig.~\ref{traj3}. In the unbroken region there are three real eigenvalues and
the system exhibits three single CPA. An example of $\Theta$ in the unbroken region is
shown in Fig.~\ref{theta3} (a) and (b) with the dashed line for $\Omega=0.79$.
By further increase of $\Omega$, two eigenvalues coalesce at another EP and the system 
returns to the $\mathcal{PT}$-broken phase. 
At the two transition points separating the broken and unbroken phases, shown by the 
vertical lines in Fig.~\ref{traj3}, the system acquires one EPs and one real eigenvalue.
This corresponds to a double CPA and one single CPA, which are shown in 
Fig.~\ref{theta3} (b) with the blue dotted curve.
Furthermore, an example featuring a higher order EP is shown in Fig.~\ref{traj3} by the 
dashed lines, and it is found for the same parameters as for the solid lines,
but with $\gamma_3=0.0157$. The parameter $\Theta$ at the triple EP is plotted
with the solid (red) line in Fig.~\ref{theta3} (b).
Around the triple EP the parameter $\Theta$ can be approximated as
\begin{eqnarray}
\Theta\approx \frac{|(\omega-\omega_{EP})^6(\omega+\omega_{EP})^6|}{|f_H(\omega_{EP})|^2},
\label{tripleEP}
\end{eqnarray}
revealing the fact that the higher order EP displays flattened absorption with respect
to the simple EP.

The significance of  Eqs.~(\ref{thetaapprox}) and (\ref{tripleEP}) and the quintessence of this section is that:
broadband flat perfect absorption can be achieved, 
by the fine tuning of the system parameters such as to obtain an EP  and a large value of $|f_H|$
at the EP.

\section{Conclusions}
Our results clearly display how harnessing the non-Hermiticity of acoustic metamaterials
provides a promising pathway towards improved acoustic absorbers.
Operating around the EPs of non-Hermitinan matrices, associated with 
the scattering process,  can lead to CPA through the avoided crossing of the corresponding 
eigenvalues. Enforcing  $\mathcal{PT}$-symmetry on the  non-Hermitian matrices 
leads to multiple CPAs, the number of which depends on the number of subwavelength resonators. 
At the EP separating the $\mathcal{PT}$-unbroken and broken phases, we find a flattened perfect 
absorption, which is even enhanced when the EP is of higher order. 

\section*{Acknowledgments}
This work has been supported by the PROPASYM project funded by the R\'egion Pays-de la- Loire,
and by the APAMAS project funded by Le Mans Acoustique.
\appendix

\section{Coupled mode equations}
Here we derive the coupled mode equations describing the scattering of acoustic waves, 
in a cylindrical waveguide side loaded by 
subwavelength Helmholtz resonators.
For frequencies below the first cut off frequency of the waveguide,
the propagation is considered one dimensional (namely $x$ direction).
 The linearized mass and momentum conservation laws can be written as:
\begin{eqnarray}
\frac{\partial p}{\partial t}+c^2\rho_0\frac{\partial u }{\partial x}=0, \label{mass} \\
\rho_0\frac{\partial u}{\partial t}+\frac{\partial p'}{\partial x}=0, \label{momentum} 
\end{eqnarray}
where $p$ and $u$ are  the pressure and velocity fluctuations in the waveguide. Above, we have used the 
constitutive equation $p=c^2\rho$ with $\rho$ being the density of air  and $c$ the speed of sound.
In the low frequency regime the resonators can be considered as point scatterers located at $x=0$, 
and we decompose the pressure field in the waveguide as follows 
\begin{align}
&p(x,t)=\Bigg\lbrace \begin{array}{l}
p_1^+(\xi_+)+p_1^-(\xi_-) \quad x\leq 0,\\
{}\\
p_2^+(\xi_+)+p_2^-(\xi_-) \quad x\geq 0,
\end{array} \label{ansatz}
\end{align}
where  $\xi_\pm=t\mp x/c_0$, while $^+$ and $^-$ denote right- and left-going waves respectively.

\begin{figure}[tbp!]
\includegraphics[width=8cm]{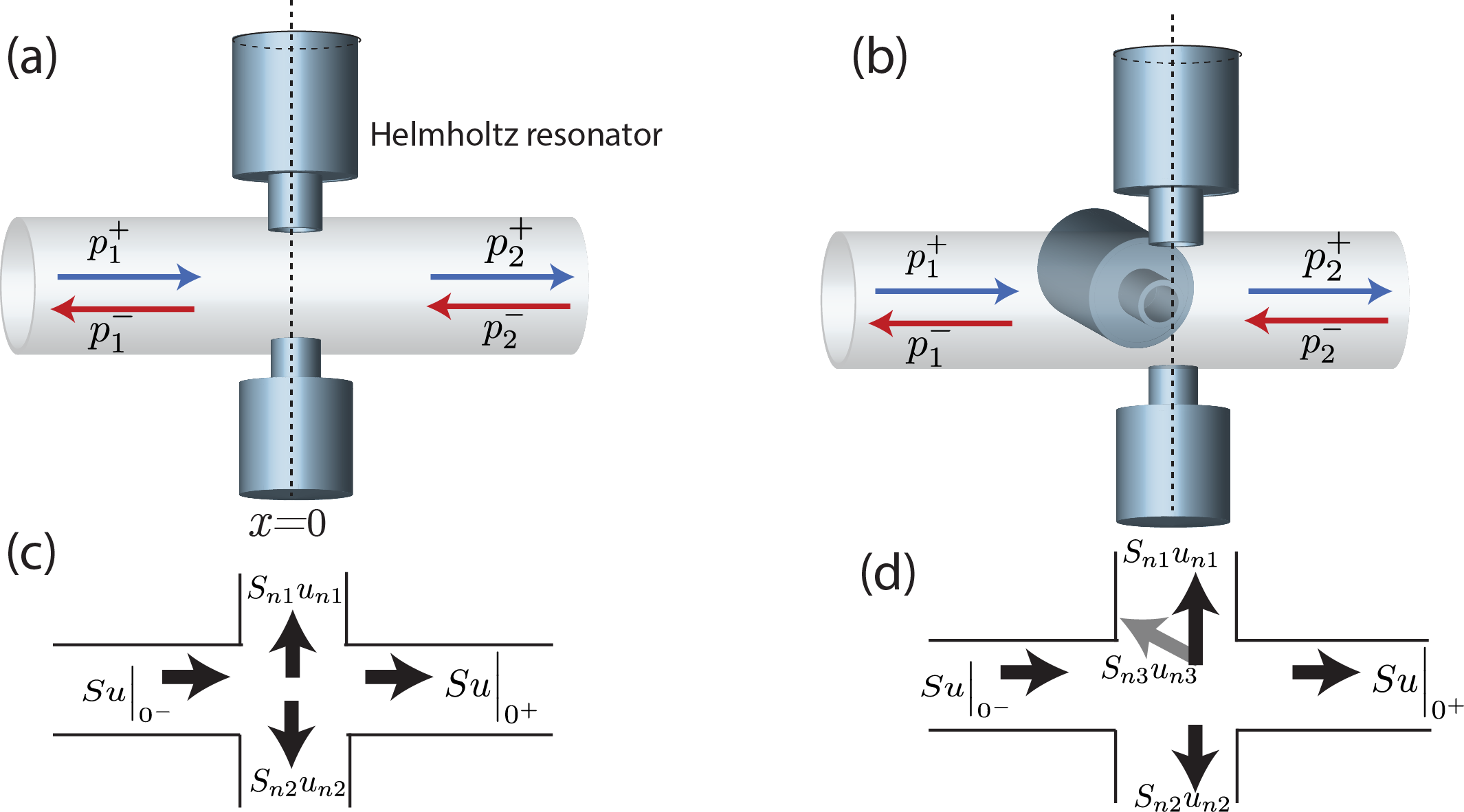}
\caption{Panels (a) and (b): depict a sketch of the configurations considered in this work
consisting of a cylindrical waveguide side loaded with two (a) or three (b) Helmholtz resonators.
Panels (c) and (d):  schematic of the continuity of acoustic flux as the waves pass through the 
point scatterers.}
\label{fig1} 
\end{figure}

The continuity of pressure at $x=0$, for both configurations is the following
\begin{align}
&p_1^{+}(t)+p_1^{-}(t)=p_2^{+}(t)+p_2^{-}t)\equiv p_0.
\label{cont1}
\end{align}
The conservation of flux at $x=0$ (illustrated in panels
(c) and (d) of Fig.~\ref{fig1}) can be written as  
 \begin{align}
S\rho_0 u\Bigr|_{x=0^-}= S\rho_0u\Bigr|_{x=0^+}+\sum_iS_{ni}\rho_0u_{ni}, \label{disc1}
 \end{align}
where $i$ takes values from $1,\ldots,{\rm N}$ for a total number of N resonators.
 $S$ and $S_{ni}$, are the cross sections of the tube and the neck of the i-th resonator respectively,
while $u_{ni}$ is the particle velocity in the neck of each Helmholtz resonator.
The velocities at the resonators neck are given by
\begin{align}
u_{ni}=-\frac{V_i}{c^2\rho_0S_{ni}}\frac{\partial p_{ci}}{\partial{t}}.
\end{align}
where $V_i$ is the volume of the cavity of the i-th resonator.
Finally it can be shown that the conservation of mass can be written as 
\begin{align}
&p_1^+(t)-p_1^-(t)=p_2^+(t)-p_2^-(t)+\sum_i\frac{V_i}{Sc_0}\frac{\partial \dot{p}_{ci}}{\partial t}.\label{cont2}
\end{align}
%
%
%
For sufficiently low frequencies, the dynamics of the pressure in each cavity of the resonators $p_{ci}(t)$,  can be described by the following approximate equations (see for example Refs.~\cite{sugi1,singh1,Kergomard1}):
\begin{align}
&\frac{\partial^2 {p}_{ci}}{\partial t^2}+\omega_i^2p_{ci}+R_{i}\frac{\partial p_{ci}}{\partial t}=\omega_i^2p_0,\\
\label{ode1}
\end{align}
where $p_0$ is the pressure at the entrance of the resonator, and  $\omega_i^2=c_0^2S_n/l_{n_i}V_i$ is the 
resonance frequency of each HR.
The resistance factor  $R_{i}$ is a parameter which quantifies the linear viscothermal losses in the resonator~\cite{singh1}.
Furthermore, using both Eqs.~(\ref{cont1})-(\ref{cont2}) we replace the pressure 
at the neck of the resonator connected with the waveguide with the following:
\begin{equation}
p_0=p_1^++p_2^--\sum_i\frac{\Gamma_i}{\omega_i^2}\dot{p}_i
\end{equation}
where $\Gamma_i=cS_{ni}/2l_{ni}S$ quantifies the coupling with the waveguide.
We use a normalized time coordinate $t\rightarrow \omega_1 t$, and the normalized parameters become
 $r_i=R_i/\omega_i$, $\gamma_i=\Gamma_i/\omega_i$ and $\Omega_i=\omega_i/\omega_1$.%

The coupled mode theory is then described by the following normalized set of equations used in the main text
\begin{align}
&p_1^{+}(t)+p_1^{-}(t)=p_2^{+}(t)+p_2^{-}(t),
\label{cont12a}\\
&p_1^+(t)-p_1^-(t)=p_2^+(t)-p_2^-(t)+\sum_i 2\frac{\gamma_i}{\Omega_i^2}\dot{p}_{ci},\label{cont22a} \\
&\ddot{\vec{P}}+\mathbf{R}\dot{\vec{P}}+\mathbf{K}\vec{P}=\vec{F},
\label{kyrillov2}
\end{align}
where $(\dot{})$ denotes differentiation with respect to the normalized time,  $\vec{P}=[p_{c1},\ldots, p_{cN}]^T$ 
and  $\vec{F}=[\Omega_1^2(p_1^++p_2^-),\ldots,\Omega_N^2(p_1^++p_2^-)]$.
The matrix $R$ has  elements
\begin{align}
R_{ii}=r_i+\gamma_i,\quad R_{ij}=\gamma_j\frac{\Omega^2_i}{\Omega^2_j}
\end{align}
while  $\mathbf{K}$ is diagonal with elements $K_{ii}=\Omega_i^2$.
Note that the above derived CMT is valid for any number of $N$ resonators,
as long as the evanescent coupling between the resonators is negligible. 

\section{Effective matrices end secular equations for the dimer and trimer}

The non-Hermitian matrix $\mathbf{L}$ for the dimer has the following form
\begin{align}
\mathbf{L}=\left( \begin{array}{cccc}
0&0&i&0 \\
0&0&0&i \\
-i&0&-i(r_{1}-\gamma_1)&-i\gamma_2/\Omega_2^2\\
0&-i\Omega_2^2 &-i\gamma_1\Omega_2^2&-i(r_{2}-\gamma_2)
\end{array}
\right).
\label{4x4_1}
\end{align}
On the other hand, for the trimer it takes the form
\begin{align}
\mathbf{L}=\left( \begin{array}{cc}
0&i\mathbf{I}_3 \\
L_{31}&L_{32}
\end{array}
\right)\label{4x4_3}
\end{align}
where $\mathbf{I}_3$ is the $3\times 3$ identity matrix, while $L_{31}$ and $L_{32}$ are given by 
the following expressions
\begin{align}
&\mathbf{L}_{31}	=\left( \begin{array}{ccc}
-i&0&0\\
0&-i\Omega_2^2&0 \\
0&0&-i\Omega_3^2\end{array}
\right),\nonumber \\
&\mathbf{L}_{32}=\left( \begin{array}{ccc}
-i(r_1-\gamma_1)&-i\gamma_2/\Omega_2^2&-i\gamma_3/\Omega_3^2\\
-i\gamma_1\Omega_2^2&-i(r_2-\gamma_2)&-i\gamma_3\Omega_2^2/\Omega_3^2 \\
-i\gamma_1\Omega_3^2&-i\gamma_2\Omega_3^2/\Omega_2^2&-i(r_3-\gamma_3)\end{array}
\right).
\label{3x3_1}
\end{align}
The corresponding characteristic polynomials are
\begin{align}
&f_{L2}=\omega^4-i\alpha\omega^3+\beta\omega^2+ i\mu\omega+\nu =0,
\label{secular1} \\
&f_{L3}=\omega^6-i\lambda'\omega^5+y'\omega^4-i\alpha'\omega^3+\beta'\omega^2+ i\mu'\omega+\nu '=0.
\label{secular31}
\end{align}
The coefficients for the dimer are the following
\begin{align}
&\alpha=r_1-\gamma1+r_2-\gamma2, \\
&\beta=\gamma_1\gamma_2\Omega_2^2-\Omega_2^2-1+(r_1-\gamma_1)(r_2-\gamma_2), \\
&\mu=\Omega_2^2(r_2-\gamma_2)+r_1-\gamma_1, \\
&\nu=\Omega_2^2,
\end{align}
and for the trimer
\begin{align}
&\lambda'=r_1-\gamma1+r_2-\gamma_2r_3-\gamma_3,  \\
&y'=-(1+\Omega_2^2+\Omega_3^2+r_1(r_2+r_3-\gamma_2\gamma_3)\nonumber \\
&+r_2(r_3-\gamma_1-\gamma_3) \nonumber -r_3(\gamma_1+\gamma2),\\
&\alpha'=r_2-\gamma_2+r_3-\gamma_3+\Omega_2^2(r_1-\gamma_1+r_3-\gamma_3)\nonumber \\
&+\Omega_3^2(r_1-\gamma_1+r_2-\gamma_1)+r_2r_3(r_1-\gamma_1)\nonumber \\
&-r_1(r_2\gamma_3+r_3\gamma_2), \\
&\mu'=\Omega_2^2(r_2-\gamma_2)+r_1-\gamma1, \\
&\nu'=\Omega_2^2.
\end{align}
\section{Exceptional points for the dimer}
\begin{figure}[tbp!]
\includegraphics[width=8cm]{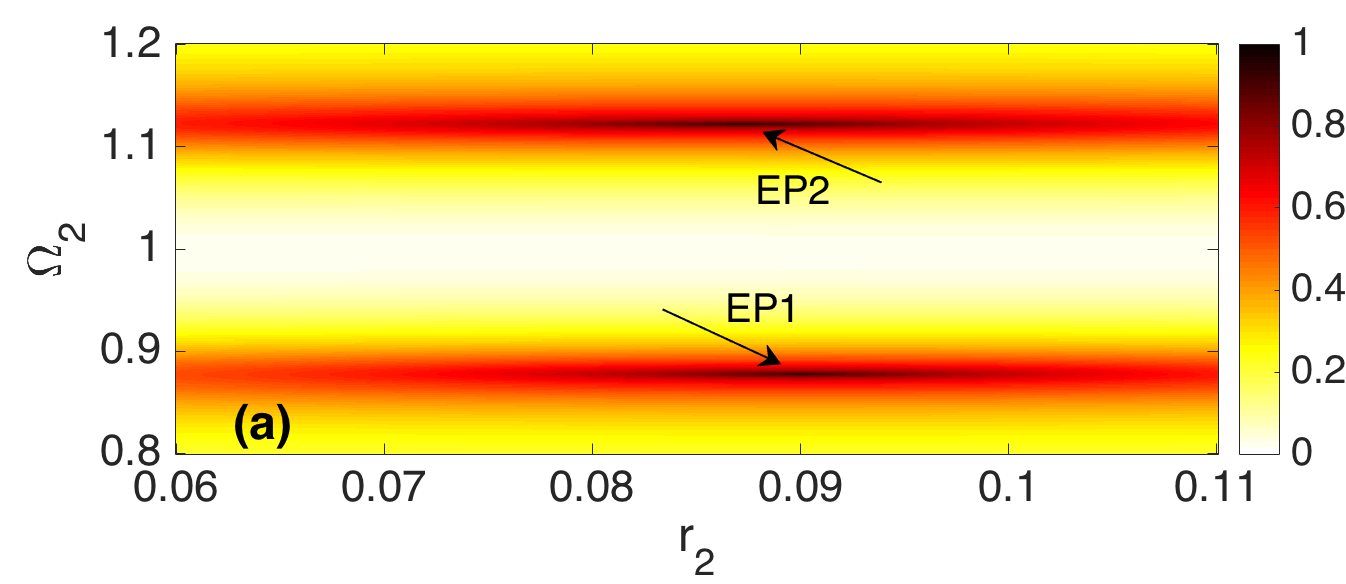}
\includegraphics[width=8cm]{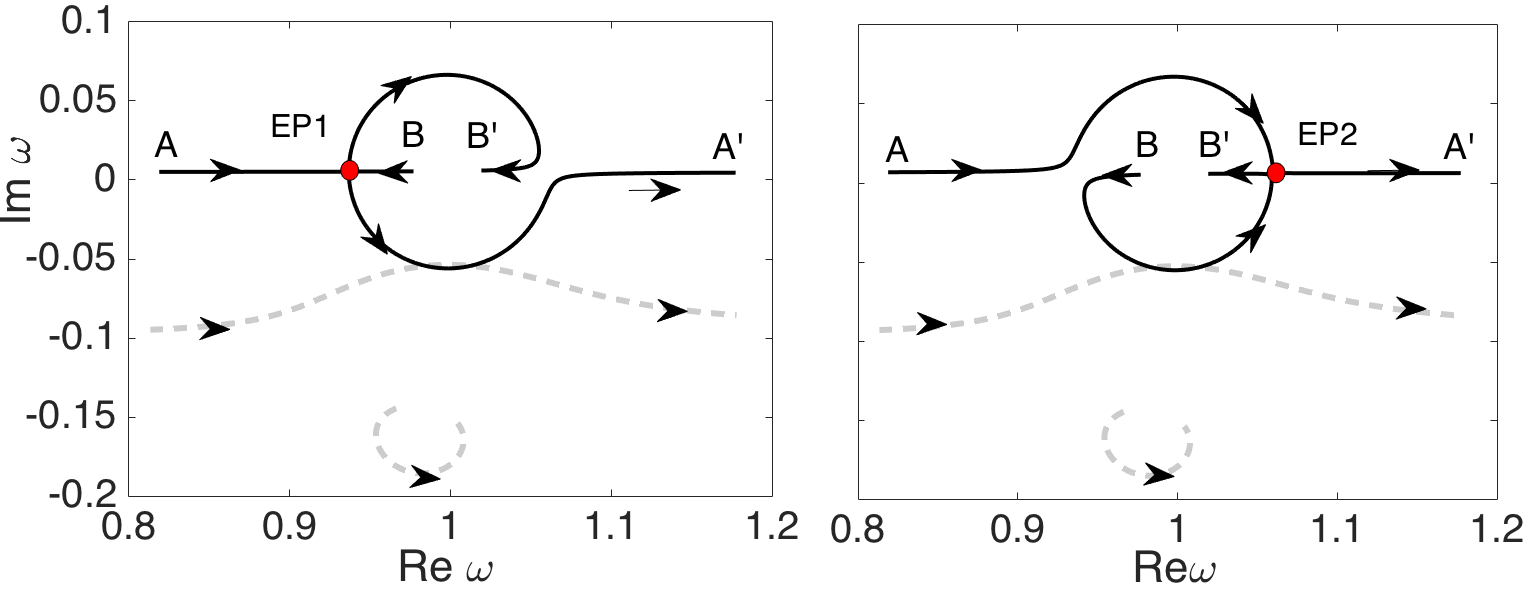}
\caption{Panel (a): A contour plot showing the parameter $|\sigma-1|$ in the $\Omega$, $r_2$ plane.
The arrows indicate the position of the two EPs. Panels (b) and (c): The eigenvalues of the matrices 
$\mathbf{L}$ (solid lines) and  $\mathbf{H}$ (dashed lines) for a trajectory with $\Omega_2\in [0.8, 1.2]$.
and $r_2=0.0872$ for panel (b) and $r_2=0.09$  for panel (c).
The rest of the parameters are $\gamma_1=0.15$, $\gamma_2=0.1$ and $r_1=0.138$.}
\label{rigit} 
\end{figure}

To locate the exceptional points of matrix $\mathbf{L}$ given by Eq.~(\ref{3x3_1})
we solve the corresponding eigenvalue problem numerically.
Since the matrix is non-Hermitian, the left and right eigenvectors
$|\phi^r\rangle$ and $|\phi^l\rangle$ are generally different,
and satisfy the following equations
\begin{align}
&\mathbf{L}|\phi^r\rangle=\omega|\phi^r\rangle,\quad \langle\phi^l|\mathbf{L}=\langle\phi^l|\omega.
\label{eig1}
\end{align}
These two vectors are biorthogonal satisfying the relation 
\begin{eqnarray}
\langle \phi^l_i|\phi^r_j\rangle=\delta_{ij},
\end{eqnarray}
and the phase rigidity $\sigma_i$ for the i-th eigenvector is defined as 
\begin{eqnarray}
\sigma=\frac{1}{\langle \phi^r|\phi^r\rangle}.
\end{eqnarray}
When the matrix is away from an exceptional point, the eigenvectors are almost orthonormal with $\sigma_i\rightarrow 1$.
On the other hand, approaching the exceptional point  the eigenvectors coalesce and $\sigma\rightarrow 0$. 
The quantity $|\sigma-1|$ is then an appropriate indicator to  identify EPs when it becomes unity.
In our case, we fix the parameters to  $\gamma_1=0.15$, $\gamma_2=0.1$ and $r_1=0.138$ 
and scan the two parameter space of $\Omega_2$ and $r_2$ to find the EPs.
For each point in this surface we calculate the quantity $|\sigma-1|$ and we plot the 
results in Fig.~\ref{rigit}. As it is shown, two EPs are found, ${\rm EP1}=\{0.8775,0.09\}$ and ${\rm EP2}=\{1.123,0.0872\}$.
Furthermore in Fig.~(\ref{rigit}) (b)-(c) we plot the eigenvalues of $\mathbf{L}$ (solid lines) and  $\mathbf{H}$
which pass through the two EPs.

\begin{thebibliography}{99}

\bibitem{moiseyev} N. Moiseyev, Non-Hermitian Quantum Mechanics
 (Cambridge University Press, London, Cambridge, 2011).

\bibitem{Heiss} W. D. Heiss, The physics of exceptional points, J. Phys. A: Math. Theor. \textbf{45} 444016 (2012).

\bibitem{Cao}   H. Cao and J. Wiersig, Dielectric microcavities: model systems for wave chaos
and non-Hermitian physics,  Rev. Mod. Phys. \textbf{87}, 61 (2015).

\bibitem{Rotter1} I. Rotter, A Non-Hermitian Hamilton Operator and the Physics of Open Quantum Systems,
 J. Phys. A \textbf{42}, 153001 (2009).

\bibitem{Bender98} C. M. Bender and S. Boettcher, Real Spectra in Non-Hermitian Hamiltonians Having PT- Symmetry, Phys. Rev. Lett. \textbf{80}, 5243 (1998).

\bibitem{Cartarius} H. Cartarius and G. Wunner, Model of a PT-symmetric Bose-Einstein condensate in a $\delta$-function double-well potential, Phys. Rev. A \textbf{80}, 013612 (2012).

\bibitem{Achilleos} V. Achilleos, P. G. Kevrekidis, D. J. Frantzeskakis, and R. Carretero-González, Dark solitons and vortices in PT-symmetric nonlinear media: From spontaneous symmetry breaking to nonlinear PT phase transitions, Phys. Rev. A \textbf{86}, 
013808 (2012).

\bibitem{Lu} X-Y. Lü, H. Jing, J-Y. Ma and Y. Wu, PT-Symmetry-Breaking Chaos in Optomechanics, 
Phys. Rev. Lett. \textbf{114}, 253601 (2015).

\bibitem{Fleury} R. Fleury,	D. Sounas and A. Alù, An invisible acoustic sensor based on parity-time symmetry,
Nat. Comm. \textbf{6}, 5905 (2015).

\bibitem{Shi} C. Shi, M. Dubois, Y. Chen, L. Cheng, H. Ramezani, Y. Wang and X. Zhang, Accessing the exceptional points of parity-time symmetric acoustics, Nat. Comm. \textbf{7}, 11110 (2016).

\bibitem{Ruter} C. E. Rüter, K. G. Makris, R. El-Ganainy, D. N. Christodoulides, M. Segev, and D. Kip, Observation of parity–time symmetry in optics, Nat. Phys. \textbf{6}, 192 (2010).


\bibitem{Ramezani} Z. Lin, H. Ramezani, T. Eichelkraut, T. Kottos, H. Cao, and D. N. Christodoulides, 
Unidirectional Invisibility Induced by PT-Symmetric Periodic Structures, Phys. Rev. Lett. \textbf{106}, 213901 (2011).

\bibitem{Regensburger} A. Regensburger, C. Bersch, M.-A. Miri, G. Onishchukov, D.N. Christodoulides, and U. Peschel,
Parity-time synthetic photonic lattices, Nature \textbf{488}, 167 (2012).

\bibitem{Peng14a} B. Peng, Ş. K. Özdemir, F. Lei, F. Monifi, M. Gianfreda, G. L. Long, S. Fan, F. Nori, C. M. Bender, and L. Yang,
Parity–time-symmetric whispering-gallery microcavities, Nat. Phys. \textbf{10}, 394 (2014).

\bibitem{Peng14b} B. Peng, Ş. K. Özdemir, S. Rotter, H. Yilmaz, M. Liertzer, F. Monifi, C. M. Bender, F. Nori, and L. Yang,
Loss-induced suppression and revival of lasing, Science \textbf{346}, 328 (2014). 

\bibitem{Chang} L. Chang, X. Jiang,S. Hua, C. Yang, J. Wen, L. Jiang, G. Li, G. Wang, and M.
Xiao,Parity–time symmetry and variable optical isolation in active–passive-coupled microresonators, Nat. Photonics \textbf{8}, 524 (2014).

\bibitem{Feng14} L. Feng, Z. J. Wong, R.-M. Ma, Y. Wang, and X. Zhang, Single-mode laser by parity-time symmetry breaking,
Science \textbf{346}, 972 (2014).

\bibitem{Hodaei} H. Hodaei, M.-A. Miri, M. Heinrich, D. N. Christodoulides,
and M. Khajavikhan, Parity-time-symmetric microring lasers, Science \textbf{346}, 975 (2014).

\bibitem{Jing} H. Jing, S. K. Özdemir, X.-Y. Lü, J. Zhang, L. Yang, and F. Nori, PT-Symmetric Phonon Laser, Phys. Rev. Lett. \textbf{113}, 053604 (2014).


\bibitem{Mostafazadeh} A. Mostafazadeh, Invisibility and PT symmetry, Phys. Rev. A \textbf{87}, 012103 (2013).
 

\bibitem{Feng12} L. Feng, Y-L. Xu, W. S. Fegadolli, M-H. Lu, J. E. B. Oliveira, V. R. Almeida, Y-F. Chen and A. Scherer, Experimental demonstration of a unidirectional reflectionless parity-time metamaterial at optical frequencies, Nat. Mat.\textbf{12}, 108 (2013).

\bibitem{Ramezani16} H. Ramezani, Y. Wang, E. Yablonovitch, X.Zhang, Unidirectional Perfect Absorber, IEEE Journal of Selected Topics in Quantum Electronics, \textbf{22}, 5 (2016). 

\bibitem{Longhi} S. Longhi, PT-symmetric laser absorber, Phys. Rev. A \textbf{82}, 031801(R) (2010)

\bibitem{Chong11} Y. D. Chong, L. Ge and A. D. Stone, PT -Symmetry Breaking and Laser-Absorber Modes in Optical Scattering Systems, Phys. Rev. Lett. \textbf{106}, 093902 (2011).

\bibitem{Chong10} Y. D. Chong, L. Ge, H. Cao, and A. D. Stone, Coherent Perfect Absorbers: Time-Reversed Lasers, Phys. Rev. Lett. \textbf{105}, 053901 (2010).

\bibitem{Veronis1} Y. Huang,  C. Min,  and G. Veronis, "Broadband near total light absorption in non-PT-symmetric waveguide-cavity systems, Opt. Express \textbf{24}, 22219 (2016).

\bibitem{Rotterpnas} B. Peng, Ş. K. Özdemir, M. Liertzer, W. Chen, J. Kramer, H. Yılmaz, J. Wiersig, 
S. Rotter, and L. Yang, Chiral modes and directional lasing at exceptional points, PNAS \textbf{113}
6845 (2016).

\bibitem{Rotternature} J. Doppler, A. A. Mailybaev, J. Böhm, U. Kuhl, A. Girschik,	F. Libisch,
 T. J. Milburn, P. Rabl,	N. Moiseyev and S. Rotter, Dynamically encircling an exceptional point for asymmetric mode switching,
 Nature \textbf{537} 76 (2016).
 
\bibitem{Sun} Y. Sun, W. Tan, H. Li, J. Li, and H. Chen, Experimental Demonstration of a Coherent Perfect Absorber 
with PT Phase Transition, Phys. Rev. Lett., \textbf{112}, 143903 (2014).

\bibitem{Zanotto14} S. Zanotto, F-P. Mezzapesa, F. Bianco, G. Biasiol, L. Baldacci, M.S. Vitiello, L. Sorba, R. Colombelli and A. Tredicucci, Perfect energy-feeding into strongly coupled systems and interferometric control of polariton absorption,
 Nat. Phys. (2014).

\bibitem{Zanotto16} S. Zanotto and A. Tredicucci, Universal lineshapes at the crossover between weak and strong critical coupling in Fano-resonant coupled oscillators, Sci. Rep. \textbf{6}, 24592 (2016).

\bibitem{Xu00} Y. Xu, Y. Li, R. K. Lee, and A. Yariv, Scattering-theory analysis of waveguide-resonator coupling, Phys. Rev. E \textbf{62}, 7389 (2000).

\bibitem{Ma} G. Ma, M. Yang, S. Xiao, Z. Yang, and P. Sheng,Acoustic metasurface with hybrid resonances, Nat. Mater. \textbf{13}, 873 (2014).

\bibitem{Wei} P. Wei, C. Croenne, S. Chu, and J. Li, Symmetrical and anti-symmetrical coherent perfect absorption for acoustic waves, Appl. Phys. Lett. \textbf{104}, 121902 (2014).

\bibitem{Romerosr} V. Romero-Garccıa, G. Theocharis, O. Richoux, A. Merkel, V. Tournat,
and V. Pagneux, Perfect and broadband acoustic absorption by critically coupled sub-wavelength resonators, Sci. Rep. \textbf{6}, 19519 (2016).

\bibitem{Groby1}  J.-P. Groby, W. Huang, A. Lardeau, and Y. Auregan, The use of slow sound to design simple sound absorbing materials, J. Appl. Phys. \textbf{117}, 124903 (2015).

\bibitem{Groby2} J.-P. Groby, R. Pommier, and Y. Auregan, Use of slow sound to design perfect and broadband sound absorbing materials,  J. Acoust. Soc. Am. \textbf{139}, 1660 (2016).

\bibitem{Merkel} A. Merkel, G. Theocharis, O. Richoux, V. Romero-Garcia, and V. Pagneux, Control of acoustic absorption in 1D scattering by indirect coupled resonant scatterers, Appl. Phys. Lett. \textbf{107}, 244102 (2015).

\bibitem{Romero16} V. Romero-Garcia, G. Theocharis, O. Richoux and V. Pagneux, Use of complex frequency plane to design broadband and sub-wavelength absorbers, J. Acoust. Soc. Am. \textbf{139}, 3395 (2016).

\bibitem{Noe} N. Jiménez, W. Huang, V. Romero-García, V. Pagneux, J-P Groby,
Ultra-thin metamaterial for perfect and quasi-omnidirectional sound absorption, Appl. Phys. Lett. \textbf{109},
121902 (2016).

\bibitem{Achilleos16} V. Achilleos, O. Richoux and G. Theocharis, Coherent perfect absorption induced by the nonlinearity of a Helmholtz resonator, J. Acoust. Soc. Am. \textbf{140}, 94 (2016).


\bibitem{PRXacoustics} K. Ding, G. Ma, M. Xiao, Z. Q. Zhang, and C. T. Chan,
Emergence, Coalescence, and Topological Properties of Multiple Exceptional Points and Their Experimental Realization,
Phys. Rev. X \textbf{6}, 021007 (2016).

\bibitem{nonhermitianacoustic}
Yi-Fan Zhu, Xue-Feng Zhu, Xu-Dong Fan, Bin Liang, Xin-Ye Zou, Jing Yang, Jian-Chun Cheng,
Non-Hermitian acoustic metamaterial for the complete control of sound by accessing the exceptional points,
arXiv:1605.04765.

\bibitem{bi} L. Xiong, W. Bi and Y. Aurégan, 
Fano resonance scatterings in waveguides with impedance boundary conditions, 
J. Acoust. Soc. Am. \textbf{139}, 764 (2016).

\bibitem{Bender2010} C. M. Bender, P. D. Mannheim, PT- symmetry and necessary and sufficient conditions for the reality of energy eigenvalues, Phys. Lett. A \textbf{374}, 1616 (2010).


%
%
%
%
%








%








 
%
%
%
%
%









%
%
%
%
%
%
%
%
%
%
%
%
%
%
%
%
%
%
%
%
%
%


\bibitem{sugi1} N. Sugimoto, M. Masuda, and T. Hashiguchi,"Frequency response of nonlinear oscillations of air column in a tube with an array of Helmholtz resonators",  J. Acoust. Soc. Am. \textbf{114}(4), 1772 (2003).

\bibitem{singh1} D. K. Singh, S. W. Rienstra, "Nonlinear asymptotic impedance model for a Helmholtz resonator liner,  J. Sound and Vibration \textbf{333} (15), 3536 (2014).

\bibitem{Kergomard1}
J. Kergomard and A. Garcia,  "Simple discontinuities in acoustic waveguides at low frequencies: Critical analysis and formulae",
 J. Sound and Vibration\textbf{114}(3), 465 (1987).
\end{thebibliography}
\end{document}